\documentclass[journal=ancac3,manuscript=layout]{achemso}
\setkeys{acs}{usetitle = true}

\usepackage[version=3]{mhchem} % Formula subscripts using \ce{}

\author{Audrey Sanchot}
\affiliation[]{CEMES, UPR 8011, CNRS-Universite de Toulouse, 29 rue Jeanne Marvig, BP 94347, F-31055 Toulouse, France}
\author{Guillaume Baffou}
\affiliation[]{ICFO-Institut de Ciencies Fotoniques, Mediterranean Technology Park, 08860, Castelldefels (Barcelona) Spain}
\alsoaffiliation{Present address: Institut Fresnel, CNRS, Aix-Marseille Universite, Ecole Centrale de Marseille, Domaine Universitaire Saint-Jerome, F-13397 Marseille, France}
\author{Renaud Marty}
\affiliation[]{CEMES, UPR 8011, CNRS-Universite de Toulouse, 29 rue Jeanne Marvig, BP 94347, F-31055 Toulouse, France}
\author{Arnaud Arbouet}
\affiliation[]{CEMES, UPR 8011, CNRS-Universite de Toulouse, 29 rue Jeanne Marvig, BP 94347, F-31055 Toulouse, France}
\author{Romain Quidant}
\email{romain.quidant@cemes.fr}
\phone{+349 3553 4076}
\affiliation[]{ICFO-Institut de Ciencies Fotoniques, Mediterranean Technology Park, 08860, Castelldefels (Barcelona) Spain}
\alsoaffiliation{ICREA-Institucio Catalana de Recerca i Estudis Avancats, 08010 Barcelona, Spain}
\author{Christian Girard}
\affiliation[]{CEMES, UPR 8011, CNRS-Universite de Toulouse, 29 rue Jeanne Marvig, BP 94347, F-31055 Toulouse, France}
\author{Erik Dujardin}
\email{erik.dujardin@cemes.fr}
\phone{+335 6225 7838}
\affiliation[]{CEMES, UPR 8011, CNRS-Universite de Toulouse, 29 rue Jeanne Marvig, BP 94347, F-31055 Toulouse, France}
 
\title{Plasmonic Nanoparticle Networks for Light and Heat Concentration}
\begin{document}

\begin{tocentry}
\includegraphics{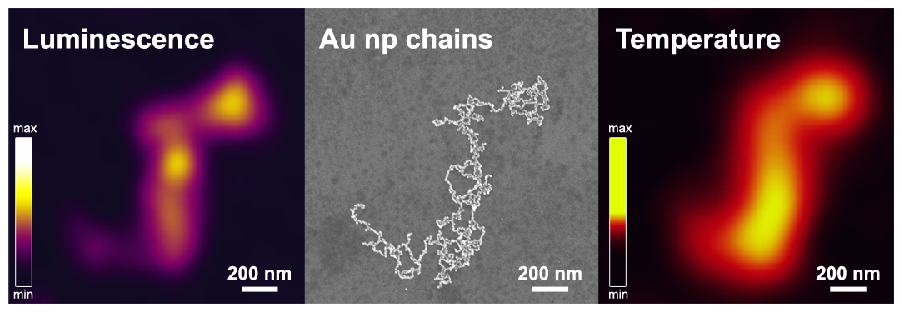} 
\end{tocentry}

\begin{abstract}
Self-assembled Plasmonic Nanoparticle Networks (PNN) composed of chains of 12-nm diameter crystalline gold nanoparticles exhibit a longitudinally coupled plasmon mode centered at 700 nm. We have exploited this longitudinal absorption band to efficiently confine light fields and concentrate heat sources in the close vicinity of these plasmonic chain networks. The mapping of the two phenomena on the same superstructures was performed by combining two-photon luminescence (TPL) and fluorescence polarization anisotropy (FPA) imaging techniques. Besides the light and heat concentration, we show experimentally that the planar spatial distribution of optical field intensity can be simply modulated by controlling the linear polarization of the incident optical excitation. On the contrary, the heat production, which is obtained here by exciting the structures within the optically transparent window of biological tissues, is evenly spread over the entire PNN. This contrasts with the usual case of localized heating in continuous nanowires, thus opening opportunities for these networks in light-induced hyperthermia applications. Furthermore, we propose a unified theoretical framework to account for both the non-linear optical and thermal near-fields around PNN. The associated numerical simulations, based on a Green s function formalism, are in excellent agreement with the experimental images. This formalism therefore provides a versatile tool for the accurate engineering of optical and thermodynamical properties of complex plasmonic colloidal architectures. 
\end{abstract}

{\bf Keywords :} self-assembled plasmonics, two-photon luminescence, light-induced heating, gold nanoparticles.

%%%%%%%%%%%%%%%%%%%%%%%%%%%%%%%%%%%%%%%%%%%%%%%%%%%%%%%%%%%%%%%%%%%%%
%% Start the main part of the manuscript here.
%%%%%%%%%%%%%%%%%%%%%%%%%%%%%%%%%%%%%%%%%%%%%%%%%%%%%%%%%%%%%%%%%%%%%

%\section{Introduction}

The optical properties of suspensions of individual colloidal metal nanoparticles have been scrutinized for more than a century, yet the rational design of their spectral properties by
 the tailoring of their size and shape has only become a topic of interest recently \cite{Tao2008}. 1 In the general context of plasmon-based technology, the next challenge for 
 nanoparticle-based plasmonics is the control of the coupling between localized surface plasmons (LSP) by mastering the spatial organization of nanoparticles \cite{Kinge2008, 
 Lin2005a,Zhang2006}. Self-assembly and templating  principles have been proposed as the most adequate approaches to promote interparticle coupling and ordered multi-scale
  organization towards the design of specific nano-optical functionalities.

\begin{figure}[hhh]
\centering\includegraphics[width=12cm,angle =0.]{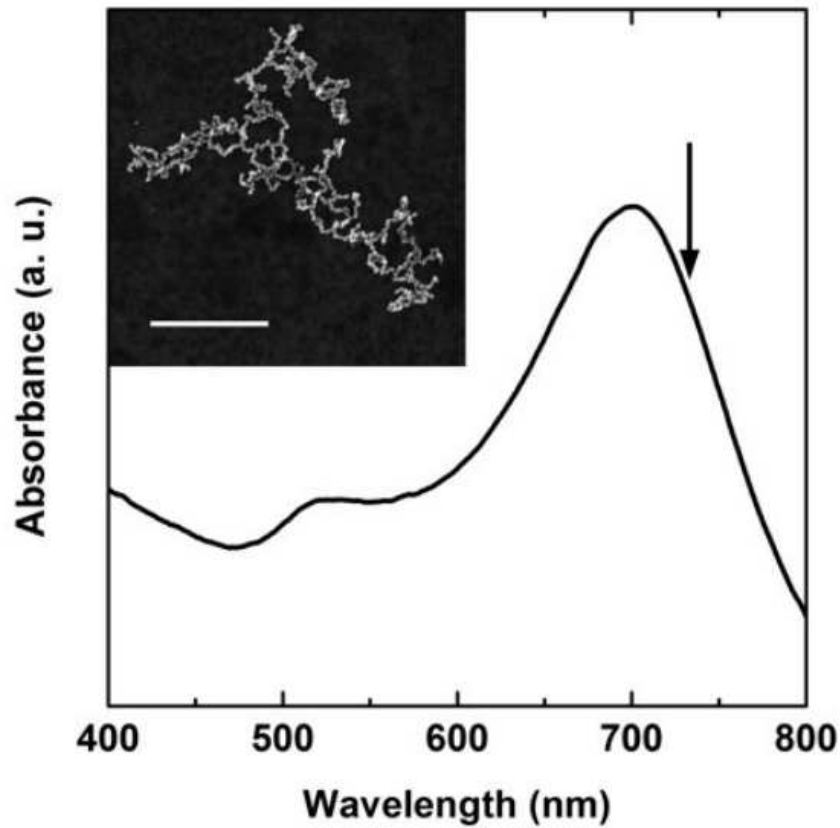}
\caption{Typical extinction spectrum of plasmonic nanoparticle networks (PNN) revealing both transversal (520 nm) and longitudinal (~700 nm) absorption bands. The arrow 
indicates the laser excitation of the longitudinal absorption band in our experiments. Inset: SEM image of a PNN. Scale bar: 500 nm. }
\label{Figure1}
\end{figure}

In this context, we have recently reported the simple fabrication of complex  and extended networks of interconnected chains of gold nanoparticles by a spontaneous  one-pot self-
assembly process driven by interparticle dipolar interactions \cite{Lin2005a, Zhang2008, Sardar2008}. In suspension, these superstructures exhibit an overall globular size of 
typically 2-3 $\mu$m but individual chain segments are one-nanoparticle, {\it i.e.} 12 nm, wide. As illustrated in \ref{Figure1}, the optical absorption spectra of these superstructures, 
that we call Plasmonic Nanoparticle Networks (PNN), display not only a transverse plasmon mode (520 nm) but also a lower energy mode (700 nm) resulting from the strong 
coupling of nanoparticle surface plasmons between neighbouring nanoparticles along the chains and that we therefore called the PNN longitudinal mode \cite
{Girard2006a,Girard2006} .

More recently, suspended PNN were deposited as intact and well-spread  networks  by deve\-loping a specific substrate surface chemistry in order to control deposition and drying
 regime of the liquid colloidal solution \cite{Bonell2009}.

The combination of the micrometer-scale size of particle assembly and the 12-nm feature size of individual colloidal constituents confers to these self-assembled architectures the 
unique ability to modulate very finely the spatial distribution of the electromagnetic field intensity. Once brought onto a solid surface, these colloidal superstructures could contribute 
to the improvement of applications as diverse as sensors, optical interconnects, enhanced nanoscale spectroscopy and microscopy.

Concomitantly, the optical characterization down to sub-wavelength and molecular-scale features has made significant progress with the recent development of near-field \cite{Girard2006a, Dickson2000, Imura2004, Imura2004a} and single molecule microscopy \cite{Gerton2004,Frey2004}, electron energy loss spectroscopy (EELS) and imaging \cite{Nelayah2007} or even suitable far- field microscopy methods \cite{Arbouet2004,Betzig2006,Rust2006,Gaiduk2010,Celebrano2011}. In particular, two-photon luminescence (TPL), has been implemented in both near-field \cite{Imura2004, Imura2004a} and far-field \cite{Bouhelier2003,Bouhelier2005,Ghenuche2008} configurations. In the latter case, the TPL intensity can be related, in first approximation, to the square of the field intensity in the metal, {\it i.e.} the fourth power of the local electric field \cite{Ghenuche2008}. With such a fast-varying spatial dependency, far-field TPL microscopy reaches a spatial resolution as high as 200 nm \cite{Baffou2010b}. Interestingly, optically-induced dissipation in plasmonic structures has also been characterized locally by far-field microscopy techniques such as Fluorescence Polarization Anisotropy imaging \cite{Baffou2010b}. This indirect temperature measurement consists in measuring the loss of fluorescence anisotropy of a molecular probe in solution, upon linearly polarized excitation, when it crosses regions of elevated temperature due to heat dissipation in the vicinity of the illuminated plasmonic structure. When the entire plasmonic structure is illuminated while the FPA is probed locally, one measures the temperature distribution as detailed in references \cite{Baffou2009a,Baffou2010}. On the contrary, when the illumination of the metal is confocal to the fluorescence excitation, then the FPA signal is related to the local heat source density \cite{Baffou2010b}.

%%%%%%%%%%%%%%%%%%%%%%%%%%%%%%%%%%%%%%%%%%%%%%%%%%%%%%%%%%
%%%%%%%%%%%%%%%%%%%%%%%%%%%%%%%%%%%%%%%%%%%%%%%%%%%%%%%%%%%

In this article, we report on the experimental observations and numerical simulations of near-field optical intensity and temperature mapping in the vicinity of gold PNNs. The large longitudinal band of these networks (Fig.\plainref{Figure1}) is used to excite the coupled modes in the chains around 730 nm with a hybrid microscope combining Two-Photon Luminescence (TPL) and Fluorescence Polarisation Anisotropy (FPA) measurements.

%%%%%%%%%%%%%%%%%%%%%%%%%%%%%%%%%%%%%%%%%%%%%%%%%%%%%%%%%%%
%%%%%%%%%%%%%%%%%%%%%%%%%%%%%%%%%%%%%%%%%%%%%%%%%%%%%%%%%%%
%%%%%%%%%%%%%%%%%%%%%%%%%%%%%%%%%%%%%%%%%%%%%%%%%%%%%%%%%%%

%\section{Methods}
%%%%%%%%%%%%%%%%%%%%%%%%%%%%%%%%%%%%%%%%%%%%%%%%%%%%%%%%%%%
%%%%%%%%%%%%%%%%%%%%%%%%%%%%%%%%%%%%%%%%%%%%%%%%%%%%%%%%%%%
%%%%%%%%%%%%%%%%%%%%%%%%%%%%%%%%%%%%%%%%%%%%%%%%%%%%%%%%%%%

\begin{figure}[hhh]
\centering\includegraphics[width=10.cm,angle =0.]{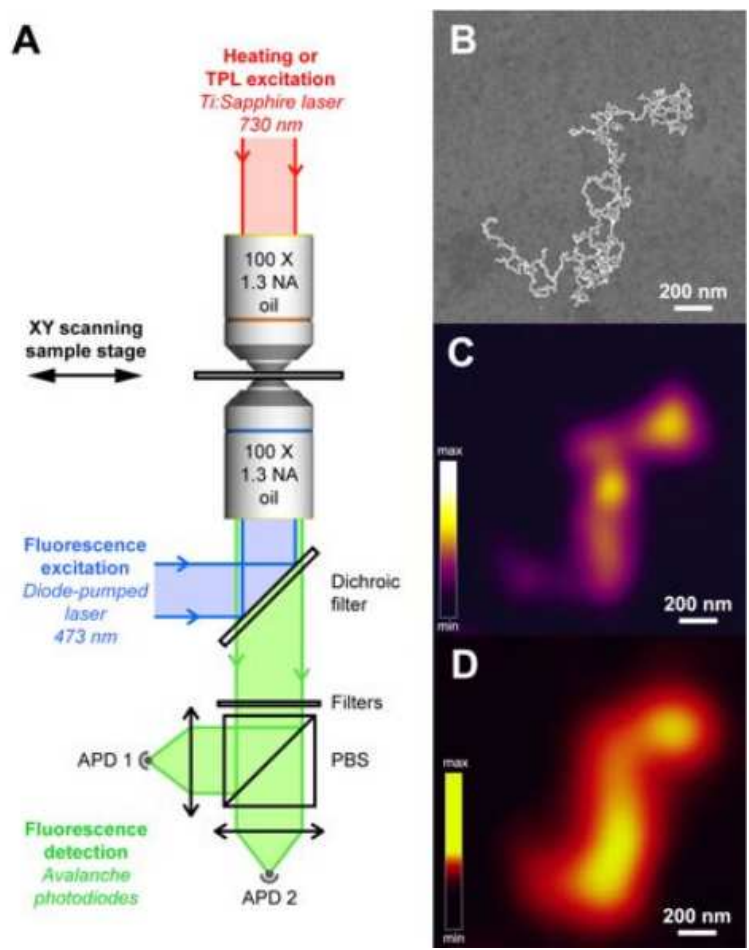}
\caption{(Color online). (A) Schematic view of the combined TPL-FPA experimental set-up. (B) SEM image of a small PNN deposited on ITO/glass substrate. (C) TPL and (D) FPA 
images of the same PNN excited at 730 nm.}
\label{Figure2}
\end{figure}

The investigation of the plasmonic properties of lithographically designed metal structures by this combined technique was first introduced in reference \cite{Baffou2010b} and the
 experimental  set-up is illustrated in \ref{Figure2}A. Briefly, isometric gold nanoparticles were freshly prepared by the citrate reduction method \cite{Frens1973} at a citrate:[AuCl$_{4}$]$^{-}$ molar ratio of 5.2:1 and diluted to the required concentration with 18 M$\Omega$ deionized water. The average diameter of the Au nanoparticles used in all experiments was 11.8 nm $^{+}_{-}$ 1.1 nm. The assembly of the PNN was performed at room temperature by adding 2-mercaptoethanol (MEA; HS(CH$_{2}$)$_{2}$OH) to Au nanoparticle solutions at a Au nanoparticle:MEA molar ratio of 1:5000.The nanoparticle chain assembly is characterized by a color change from pink to purple as the coupled mode emerges and was monitored by UV-vis spectrophotometry. The assembly is complete within 24 to 48 hours after mixing.
 %%%%%%%%%%%%%%%%%%%%%%%%%%%%%%%%%%%%%%%%%%%%%%%%%%%%%%%%%%%
%%%%%%%%%%%%%%%%%%%%%%%%%%%%%%%%%%%%%%%%%%%%%%%%%%%%%%%%%%%
 Freshly prepared PNN are deposited on cleaned glass coverslips  coated with 10 nm conductive Indium Tin Oxide (ITO) layer \cite{Bonell2009}. Well-spread networks were identified by FEG-SEM imaging (Fig.\plainref{Figure2}B). To perform the TPL measurements, an IR Ti:Sapphire laser operated in pulsed mode is tightly focused on the sample (Fig.\plainref{Figure2}B) with a high numerical aperture objective (NA=1.3). The typical IR laser power is 1-3 mW and the pulse duration is 180 fs. In all the experiments, the IR laser wavelength was set to 730 nm. An image is obtained by raster scanning the sample in X and Y using a piezo stage and collecting the emitted light on an avalanche photodiode through a NA=1.3 objective (Fig.\plainref{Figure2}C).

To perform the temperature increase measurements, two laser beams are focused and overlapped \cite{Baffou2009}. The IR Ti:Sapphire laser is set in cw mode to locally heat the
 PNN, while a second linearly polarized blue laser is used to excite the fluorescein molecules dispersed in a glycerol/water mixture surrounding the structure. The typical power of 
 the blue and IR laser are 60 $\mu$W and 4 mW respectively. The emitted molecular fluorescence is collected through the bottom objective and sent to two APDs {\it via} a polarizing 
 cubic beam splitter to construct two fluorescence maps corresponding to 2 perpendicular emission polarization directions. The loss in fluorescence anisotropy is calculated from 
 the polarization difference signal and the associated temperature map is then obtained by using a FPA = f(T) calibration curve described elsewhere (Fig.\plainref{Figure2}D) \cite
 {Baffou2009}. To get the temperature increase as function of the IR laser location, the sample is raster scanned in X and Y as for the TPL measurements. For both types of experiments, particular care was taken to strictly avoid any damage of the nanoparticles by the laser-induced heating. The onset of melting could easily be identified in TPL by the rapid variation and eventually disappearance of the TPL signal within a few pixels. It usually occurred for nominal IR laser power of about 10 mW. The IR laser power used in all experiments was kept below 30 \% of that needed for damaging the nanoparticles.

Finally, the SEM images of the measured samples were resolved enough to extract positional maps of each individual 12-nm diameter nanoparticle from the studied PNN. These
 experimental maps were used to compute the simulated TPL and FPA maps. 
 
This theoretical analysis was based on a novel numerical tool involving the computation of temperature increase and TPL signal through a complete self-consistent scheme using 
both electrodynamical and thermal Green functions \cite{Baffou2010c}. In the following, our model is first described in detail and experimental images are then compared to 
simulated ones for both TPL and FPA signals of one in several studied PNN.

%%%%%%%%%%%%%%%%%%%%%%%%%%%%%%%%%%%%%%%%%%%%%%%%%%%%%%%%%%%%
%%%%%%%%%%%%%%%%%%%%%%%%%%%%%%%%%%%%%%%%%%%%%%%%%%%%%%%%%%%%
%%%%%%%%%%%%%%%%%%%%%%%%%%%%%%%%%%%%%%%%%%%%%%%%%%%%%%%%%%%%

%\section{Results and Discussion}

%%%%%%%%%%%%%%%%%%%%%%%%%%%%%%%%%%%%%%%%%%%%%%%%%%%%%%%%%%%
%%%%%%%%%%%%%%%%%%%%%%%%%%%%%%%%%%%%%%%%%%%%%%%%%%%%%%%%%%%
%%%%%%%%%%%%%%%%%%%%%%%%%%%%%%%%%%%%%%%%%%%%%%%%%%%%%%%%%%%

\begin{figure}[hhh]
\centering\includegraphics[width=12.cm,angle =0.]{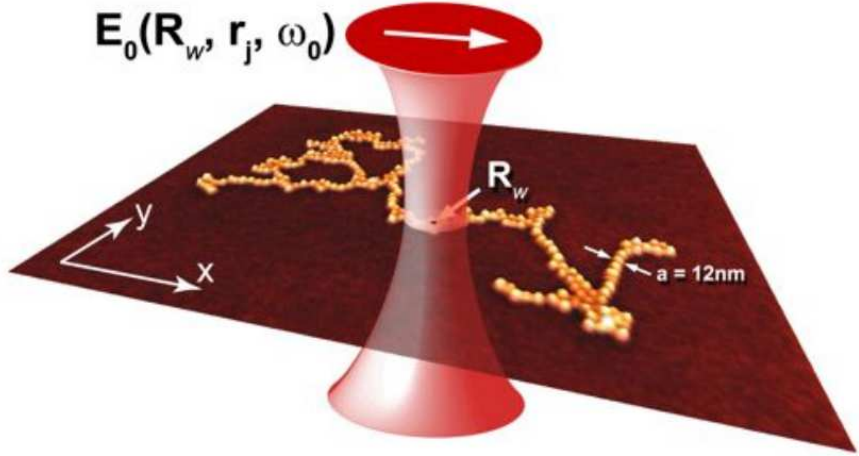}
\caption{(color online) Schematic view of the experimental configuration in which a PNN made of 12-nm diameter nanoparticles is illuminated in normal incidence by a focus laser
 beam characterized by an electric field vector ${\bf E}_{0}({\bf R}_w,{\bf r}_{i},\omega_{0})$, ${\bf R}_{w}$, ${\bf r}_{i}$ and $\omega_{0}$ being the beam waist position, the 
 considered location for ${\bf E}_{0}$ and the incident field pulsation.}
\label{Figure3}
\end{figure}

At the nanometer scale, the optical properties of small plasmonic objects deposited on a surface (small individual or self-assembled colloidal particles) can be investigated by 
adapting the coupled dipole approximation (CDA) method \cite{DRAINE1994} to a planar geometry \cite{Girard2006a}. We first describe the unified formalism and numerical 
method that we developed in this context to compute the TPL and temperature distributions near PNNs, as depicted in \ref{Figure3}. 
%%%%%%%%%%%%%%%%%%%%%%%%%%%%%%%%%%%%%%%%%%%%%%%%%%%%%%%%%%%
%%%%%%%%%%%%%%%%%%%%%%%%%%%%%%%%%%%%%%%%%%%%%%%%%%%%%%%%%%%
 %%%%%%%%%%%%%%%%%%%%%%%%%%%%%%%%%%%%%%%%%%%%%%%%%%%%%%%%%%%
%%%%%%%%%%%%%%%%%%%%%%%%%%%%%%%%%%%%%%%%%%%%%%%%%%%%%%%%%%%
%%%%%%%%%%%%%%%%%%%%%%%%%%%%%%%%%%%%%%%%%%%%%%%%%%%%%%%%%%%
 %%%%%%%%%%%%%%%%%%%%%%%%%%%%%%%%%%%%%%%%%%%%%%%%%%%%%%%%%%%
The network is composed of N identical 
subwavelength-sized nanoparticles (12 nm) located at the positions $r_{i}$. When the sample is illuminated  by the electric field ${\bf E}_{0}({\bf R}_w,{\bf r},\omega_{0})$ 
associated with a monochromatic laser beam of frequency $\omega_{0}$ centered and focused at ${\bf R}_w$, each nanometric component acquires an oscillating  dipole moment 
driven by the local electric field ${\bf E}({\bf R}_w,{\bf r}_{i},\omega_{0})$, which originates from the many-body interactions with the other particles:
\begin{eqnarray}
{\bf P}({\bf R}_w,{\bf r}_{i},\omega_{0})
=\alpha(\omega_{0})\cdot{\bf E}({\bf R}_w,{\bf r}_{i},\omega_{0}) 
\label{LOCAL0}
\end{eqnarray} 
where $\alpha(\omega_{0})$ characterizes the dipolar polarizability of the metal particles. The many-body interactions between the nanoparticles can be introduced by writing N 
implicit linear equations:
\begin{eqnarray}
{\bf E}({\bf R}_w,{\bf r}_{i},\omega_{0})={\bf E}_{0}({\bf R}_w,{\bf r}_{i},\omega_{0})
\nonumber
\\
+\sum_{j}{\bf S}({\bf r}_{i},{\bf r}_{j},\omega_{0})
\cdot\alpha(\omega_{0})
\cdot{\bf E}({\bf R}_w,{\bf r}_{j},\omega_{0})
\; ,
\label{LOCAL1}
\end{eqnarray}
where ${\bf S}({\bf r}_{i},{\bf r}_{j},\omega_{0})$ represents the Green dyadic function of the bare substrate \cite{Girard2006a}. 

The TPL signal arising from the gold nanoparticles is a non-coherent process characterized by a wide spectrum spanning from $\omega_0$ to $2\omega_0$. This process is related to the intensity distribution $|{\bf E}({\bf R}_w,{\bf r}_{i},\omega_{0})|^{2}$ of local electric fields experienced by each particle. To solve this problem, we assume that the TPL signal can be modeled by summing the different intensities $|{\bf E}({\bf R}_w,{\bf r}_{i},\omega_{0})|^{2}$  rather than the electric field amplitudes ${\bf E}({\bf R}_w,{\bf r}_{i},\omega_{0})$ to account for the non-coherent nature of the TPL process:
\begin{eqnarray}
I_\mathsf{TPL}(\mathbf{R}_w,\omega_0)\propto\left[\sum_{i=1}^N|\mathbf{E}_i(\mathbf{R}_w,\omega_0)|^2\right]^2
\label{ITPL-GB}
\end{eqnarray}
The overall square is intended to account for the two-photon absorption.
 
 %%%%%%%%%%%%%%%%%%%%%%%%%%%%%%%%%%%%%%%%%%%%%%%%%%%%%%%%%%%
\begin{figure}[hhh]
\centering\includegraphics[width=12.cm,angle =0.]{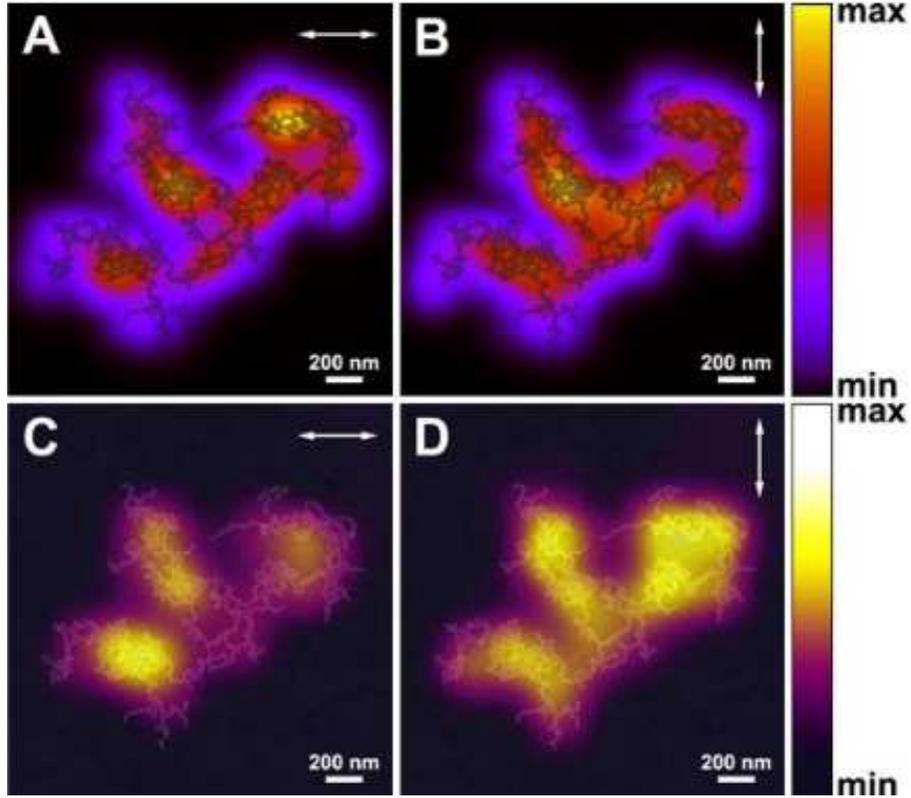}
\caption{(Color online) TPL mapping near a large PNN composed of 1537 gold nanoparticles (image size 2 x 2 $\mu$m). (A-B) Simulation images obtained from equation (\plainref
{ITPL-GB}) overlaid with the nanoparticle map. (C-D) Experimental TPL images measured on this PNN overlaid with the SEM image of the PNN. Incident polarization is aligned
along OX for (A) and (C) and along OY for (B) and (D).}
\label{Figure4}
\end{figure}
%%%%%%%%%%%%%%%%%%%%%%%%%%%%%%%%%%%%%%%%%%%%%%%%%%%%%%%%%%%

In \ref{Figure4}, we present a sequence of theoretical (Fig.\plainref{Figure4}A-B) and experimental (Fig.\plainref{Figure4}C-D) TPL images obtained by scanning the tightly focused
 Ti:sapphire laser spot with respect to a PNN composed of 1537 gold nanoparticles, the SEM image of which can be seen in the background of panels C and D. For each position of the laser
  beam waist ${\bf R}_{w}$ (incident wavelength $\lambda_{0}$ = 730 nm), the simulated TPL intensity has been computed from relation (\plainref{ITPL-GB}) by introducing the 
  coordinates of the gold particle map plotted as overlay in black circles in panels A and B into the numerical code. The two incident polarizations, investigated along $OX$ and 
  $OY$ directions, show an excellent agreement between experimental data and simulations, indicating that TPL measurements provide quantitative information on the light 
  confinement near extremely small gold structures. Although the TPL signal seems to be more intense in regions of denser particle packing, this correlation does not account for the observed polarization dependency of the images. This polarization effect suggests that the local anisotropy of uniaxially coupled nanoparticles translates into anisotropic light confinement upon excitation of the PNN at 730 nm.
  
Although limited by the intrinsic sensitivity and spatial resolution of our microscopy set-up, these observations illustrate how strongly coupled nanoparticles of small size but high 
crystallinity are able to laterally confine the electromagnetic intensity within a few tens of nanometers, while allowing its 
distribution along chains of a few hundreds of nanometers, as demonstrated theoretically elsewhere \cite{Girard2006a}.

Simultaneously, during the IR illumination, the PNN temperature rises because of Joule effect induced inside the metal particles. The heat power $Q_i$ delivered by each 
nanoparticle can be computed from the electric field amplitude $\mathbf{E}_i(\mathbf{R}_w,\omega_{0})$ using the relation\cite{Baffou2010b}:
\begin{equation}
Q_i({\bf R}_w,\omega_0)=\frac{\omega_{0}\text{Im}[\alpha(\omega_{0})]}{2}|\mathbf{E}_i({\bf R}_w,\omega_{0})|^{2}
\, .
\label{DIS-Q}
\end{equation}

From this relation, the temperature rise $\Delta T({\bf R}_{w},\omega_{0})$, expected at the laser location ${\bf R}_{w}$, can be deduced using the thermal Green's function:\cite
{Baffou2010c}
\begin{equation}
\Delta T({\bf R}_{w},\omega_{0})=\frac{1}{4\pi \kappa}
\sum_{i=1}^N\frac{Q_i({\bf R}_w,\omega_{0})}{|{\bf R}_{w}-{\bf r}_i|}
\, ,
\label{TEMP}
\end{equation}
where $\kappa$ is the thermal conductivity of the surroundings. In order to take into account the influence of the substrate conductivity, the image method has been used to compute the temperature distribution induced by the PNNs \cite{Baffou2010c}.

\begin{figure}[hhh]
\centering\includegraphics[width=12.cm,angle =0.]{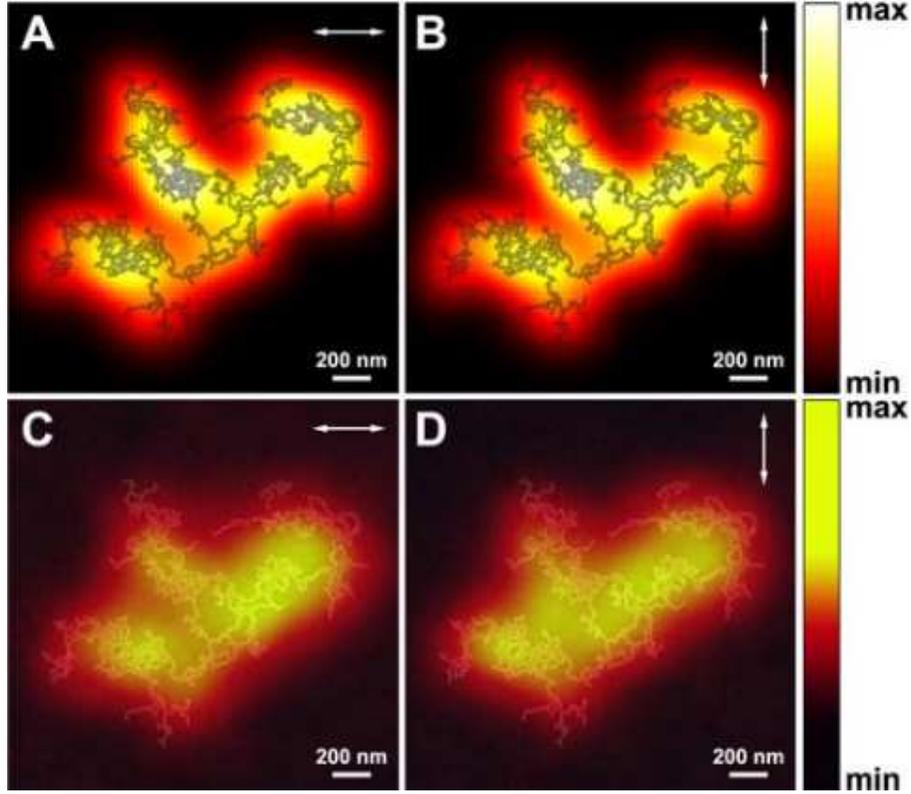}
\caption{(Color online) Temperature mapping near the same large PNN shown in \ref{Figure4}. (A-B) Simulation images obtained from \ref{TEMP}. (C-D) Experimental images 
measured by the fluorescence polarization anisotropy of fluorescein molecules surrounding the PNN. Incident polarization is aligned along $OX$ for (A) and (C) and along $OY$
 for (B) and (D). Calculated (A, B) and experimental (C, D) images are respectively overlaid with the positional network particle map and the SEM image of the PNN.}
 \label{Figure5}
\end{figure}

\ref{Figure5} shows a sequence of temperature maps resulting from the scanning of the same large PNN described in \ref{Figure4} through a 300 nm confocal spot of the near-IR 
and blue lasers . 
 %%%%%%%%%%%%%%%%%%%%%%%%%%%%%%%%%%%%%%%%%%%%%%%%%%%%%%%%%%%
%%%%%%%%%%%%%%%%%%%%%%%%%%%%%%%%%%%%%%%%%%%%%%%%%%%%%%%%%%%
Although much less confined than the optical near--field intensity revealed in the TPL maps of \ref{Figure4}, the temperature distribution faithfully reproduces the
 PNN morphology. Noticeably, these thermal maps do not display any significant variation with the incident polarization (see maps \plainref{Figure5}C and \plainref{Figure5}D). In
particular, the three main branches of the network generate a same temperature rise because they are composed of many linear chains able to support the excited longitudinal 
mode irrespectively of the incident polarization. In addition, although we deal with extremely localized heat sources, the temperature distribution appears smoother than the near-
field optical intensity map induced near the same system because of the thermal diffusion \cite{Ye2011}. Two key differences between the generation of heat and optical TPL 
intensity should be emphasized. In the first case, as illustrated by Eq. \plainref{TEMP}, the temperature distribution generated at location ${\bf R}_{obs}$ by a point-like heat source
located at ${\bf R}_{h}$ varies like the inverse of $|{\bf R}_{obs}-{\bf R}_{h}|$ while it varies like the inverse of the third power of this distance in the case of optical near-field. 
   
Furthermore, the TPL signal is sensitive to the square of the optical near-field intensity (cf. equation \plainref{ITPL-GB}), while the thermal process is just proportional to the 
global near-field intensity (cf Eq. \plainref{TEMP}). This is clearly illustrated in both experimental and theoretical images of \ref{Figure4}, where we observe a significant variation of 
the TPL pattern inside the three branches of the PNN when the polarization is rotated by 90$^{\circ}$ in contrast to the invariance of thermal maps in \ref{Figure5}.

Interestingly, the excitation of longitudinally coupled modes in PNN involves electrically separated nanoparticles in contrast to the continuous metallic stripes or nanowires prepared by 
lithographic approaches \cite{Baffou2010b} or colloidal synthesis. As a consequence, the energy dissipation originates from each individual nanoparticle and is homogeneously 
distributed along the nanoparticles chains rather than localized where the excited nanorods plasmon mode induces a maximal current as shown in reference [24]. 
The self- assembly of the nanoparticles into globular micrometer-scale PNN therefore combines a number of specificities that could be advantageously exploited in the context of 
hyperthermia cell treatment \cite{Huang2008}. Although the full assessment of the hyperthermia efficiency of PNN on live bacteria, which is currently underway, lies beyond the 
scope of this article, the assets of our system already appear in the present results. Indeed, the heat production in PNN is obtained by addressing the longitudinally coupled mode 
centered at 700 nm that allows an excitation through the transparency window of biological tissues. Moreover, by choosing this wavelength, no individual stray particles or particle 
short chains (less than 4-5 nanoparticles) is excited, therefore obviating the parasitic heat dissipation away from the PNN. Moreover, the commensurability of the PNN with cell size, 
the biocompatibility of gold nanoparticles and the spontaneous affinity of PNN for the cell membrane (Figure S1 in Supplementary Info) strongly suggest that optical energy could 
be dissipated directly at the PNN-wrapped cell walls without requiring an internalization step.

%%%%%%%%%%%%%%%%%%%%%%%%%%%%%%%%%%%%%%%%%%%%%%%%%%%%%%%%%%%
%%%%%%%%%%%%%%%%%%%%%%%%%%%%%%%%%%%%%%%%%%%%%%%%%%%%%%%%%%%
%%%%%%%%%%%%%%%%%%%%%%%%%%%%%%%%%%%%%%%%%%%%%%%%%%%%%%%%%%%
%\section{Conclusion}
%%%%%%%%%%%%%%%%%%%%%%%%%%%%%%%%%%%%%%%%%%%%%%%%%%%%%%%%%%%
%%%%%%%%%%%%%%%%%%%%%%%%%%%%%%%%%%%%%%%%%%%%%%%%%%%%%%%%%%%%
%%%%%%%%%%%%%%%%%%%%%%%%%%%%%%%%%%%%%%%%%%%%%%%%%%%%%%%%%%%

In conclusion, this work demonstrates for the first time that large colloidal microstructures made of 12 nm diameter, crystalline nanoparticles self-assembled into single particle 
chains are able to effectively modulate the 2D spatial distribution of electromagnetic intensity, once deposited onto a substrate. Since the self-assembled networks are excited in 
the longitudinal coupled mode borne by the chains, these energy distributions can be simply tuned by changing the polarization of the incident light. The investigation of the near-
field using TPL mapping provides an efficient means to characterize the intensity distribution at the scale of the entire superstructure but also reveals in-plane field modulation 
features which are only limited by the intrinsic spatial resolution of our TPL set-up, namely {\it ca.} 200 nm. Therefore, PNN are shown to have a unique potential for confining and 
enhancing the electromagnetic field down to sub-100 nm length scale.

Moreover, the measurement of the temperature distribution with our dual set-up has shown that heat is efficiently generated by exciting the longitudinal coupled mode and the 
generated heat is confined near the PNN. Understandably, the spatial distribution of the temperature elevation is more homogeneous and less sensitive to incident polarization 
than the TPL signal. Nevertheless, the optically-induced localized heat dissipation in our water-dispersible gold nanoparticle self-assembly holds promises in biomedical 
applications beyond the current capabilities of individual colloids or substrate-bound metal structures.

Finally, we have developed a model and simulation tool that accounts for most of the observed 2D field map features. This is now a useful tool for the engineering of the optical 
near-field, for example for nanoscale optical information processing devices, using self-assembled structures made of crystalline metal colloids. The extension of our model and 
simulation tool to dissipation  phenomena faithfully reproduced the observed temperature maps.

%%%%%%%%%%%%%%%%%%%%%%%%%%%%%%%%%%%%%%%%%%%%%%%%%%%%%%%%%%%
%%%%%%%%%%%%%%%%%%%%%%%%%%%%%%%%%%%%%%%%%%%%%%%%%%%%%%%%%%%
%%%%%%%%%%%%%%%%%%%%%%%%%%%%%%%%%%%%%%%%%%%%%%%%%%%%%%%%%%%

\begin{acknowledgement}
This work was supported by the European Research Council (ERC) (Grant ERC-2007-StG Nr 203872 COMOSYEL to E. D.), the French Agence Nationale de la Recherche (Grant ANR-08-NANO-054-01- NAPHO to C. G. and NT09-451197-PlasTips to E. D.), the French Ministry of Research (PhD fellowships to A. S. and R. M.), and the massively parallel computing center CALMIP in Toulouse. G. B. and R. Q. also acknowledge the financial support from Fundacio Privada CELLEX.
\end{acknowledgement}

\begin{suppinfo}
SEM images of bacteria colony and isolated individuals coated with Plasmonic Nanoparticle Networks (PNN).
\end{suppinfo}

%%%%%%%%%%%%%%%%%%%%%%%%%%%%%%%%%%%%%%%%%%%%%%%%%%%%%%%%%%%
%\bibliography{2012-ACSNano2.bib}
%%%%%%%%%%%%%%%%%%%%%%%%%%%%%%%%%%%%%%%%%%%%%%%%%%%%%%%%%%%
%%%%%%%%%%%%%%%%%%%%%%%%%%%%%%%%%%%%%%%%%%%%%%%%%%%%%%%%%%%
\providecommand*\mcitethebibliography{\thebibliography}
\csname @ifundefined\endcsname{endmcitethebibliography}
  {\let\endmcitethebibliography\endthebibliography}{}

\end{document}